\documentclass[aps,prd,showpacs,floatfix,preprint]{revtex4}
\usepackage{amsmath,bm}
\usepackage{graphicx}
\usepackage{epsfig}
\begin{document}

\title{Hyperon bulk viscosity in strong magnetic fields}

\author{ Monika Sinha and Debades Bandyopadhyay}
\affiliation{Theory Division and Centre for Astroparticle Physics,
Saha Institute of Nuclear Physics, 1/AF Bidhannagar, 
Kolkata-700064, India}

\begin{abstract}
We study the bulk viscosity of neutron star matter including $\Lambda$ hyperons 
in the presence of quantizing magnetic fields. Relaxation time and bulk 
viscosity due to both the non-leptonic weak process involving $\Lambda$ 
hyperons and direct Urca processes are calculated here. In the 
presence of a strong magnetic field of $10^{17}$ G, the hyperon bulk viscosity 
coefficient is reduced whereas bulk viscosity coefficients due to direct Urca 
processes are enhanced compared with their field free cases when many Landau 
levels are populated by protons, electrons and muons. 
\pacs{97.60.Jd, 26.60.-c, 04.40.Dg}
\end{abstract}

\maketitle

\section{Introduction}
R-mode instability plays an important role in regulating spins of newly born
neutron stars as well as old and accreting neutron stars in low mass x-ray
binaries \cite{Nar}. Gravitational radiation drives the r-mode 
unstable due to Chandrasekhar-Friedman-Schutz mechanism 
\cite{Chan,Frie,Kok,And01,And98,Fri98,Lin98,And99,Ster}. R-mode instability
could be a promising source of gravitational radiation. It would be possible to
probe neutron star interior if it is detected by gravity wave detectors.

Like gravitational radiation, electromagnetic radiation also drives the r-mode 
unstable through Chandrasekhar-Friedman-Schutz mechanism. There exists a class 
of neutron stars called magnetars \cite{Tho} with strong surface magnetic 
fields $10^{14}-10^{15}$ G as
predicted by observations on soft gamma-ray repeaters and anomalous x-ray 
pulsars \cite{Kou,Vas}. The effects of magnetic fields on the spin 
evolution and r-modes in protomagnetars were investigated by different groups 
\cite{Rez1,Lai,Rez2}. On the one hand, it was shown that the 
growth of the r-mode due to electromagnetic and Alfv\'en wave emission 
for strong magnetic field and slow rotation could compete with that of 
gravitational radiation \cite{Lai}. On the other hand, it was 
argued that the distortion of magnetic fields in neutron stars due to r-modes
might damp the mode when the field is $\sim 10^{16}$ G or more \cite{Rez1,Rez2}.

The evolution of r-modes proceeds through three steps \cite{Owen}. In the 
first phase, the mode amplitude grows exponentially with time. In the next 
stage, the mode saturates due to nonlinear effects. In this case viscosity 
becomes important. Finally, viscous forces dominate over gravitational 
radiation driven instability and damp the r-mode. This shows that viscosity 
plays an important role on the evolution of r-mode. Bulk and shear viscosities 
were extensively investigated in connection with the damping of the r-mode 
instability
\cite{Nar,Jon1,Jon2,Lin02,Dal,Dra,Chat1,Chat2,Chat3,Han,And06,Mad92,Mad00,Don1,Don2,Pan,Bas,Schm,Alf,Bas2}. In particular, it was shown that the
hyperon bulk viscosity might effectively damp the r-mode instability
\cite{Chat3}. However all these calculations of viscosity were performed in the
absence of magnetic fields. The only calculation of bulk viscosity due to Urca 
process in magnetised neutron star matter was presented in Ref.\cite{anand}. 
This motivates us to investigate bulk viscosity due to non-leptonic process 
involving hyperons in the
presence of strong magnetic fields. It is to be noted that the magnetic field
in neutron star interior might be higher by several orders of magnitude than
the surface magnetic field \cite{Lai2}. Further it was shown that neutron stars 
could sustain strong interior magnetic field $\sim 10^{18}$ G \cite{Car,Bro}.  

The paper is organised in the following way. In Section \ref{eos} we 
describe hyperon
matter in strong magnetic fields. We calculate bulk viscosity due to 
the non-leptonic process 
involving $\Lambda$ hyperons and due to leptonic processes in Section \ref{bv}.
We discuss results in Section \ref{rd} and a summary is
given in Section \ref{sum}. 

\section{\label{eos}Hyperon matter in magnetic field}
We describe $\beta$ equilibrated and charge neutral neutron star matter made of
neutrons, protons, $\Lambda$ hyperons, 
electrons and muons within a relativistic mean field approach 
\cite{walecka, serot}. The baryon-baryon interaction is mediated by 
$\sigma$, $\omega$ and $\rho$ mesons. In
the absence of magnetic field, the baryon-baryon interaction is given by 
the Lagrangian density \cite{glendenning,schaffnerprc}
\begin{eqnarray}
{\cal L}_B &=& \sum_{B=n,p,\Lambda} \bar\psi_{B}\left(i\gamma_\mu 
{\partial}^{\mu} - m_B
+ g_{\sigma B} \sigma - g_{\omega B} \gamma_\mu \omega^\mu
- g_{\rho B}
\gamma_\mu{\mbox{\boldmath t}}_B \cdot
{\mbox{\boldmath $\rho$}}^\mu \right)\psi_B\nonumber\\
&& + \frac{1}{2}\left( \partial_\mu \sigma\partial^\mu \sigma
- m_\sigma^2 \sigma^2\right) - U(\sigma) \nonumber\\
&& -\frac{1}{4} \omega_{\mu\nu}\omega^{\mu\nu}
+\frac{1}{2}m_\omega^2 \omega_\mu \omega^\mu
- \frac{1}{4}{\mbox {\boldmath $\rho$}}_{\mu\nu} \cdot
{\mbox {\boldmath $\rho$}}^{\mu\nu}
+ \frac{1}{2}m_\rho^2 {\mbox {\boldmath $\rho$}}_\mu \cdot
{\mbox {\boldmath $\rho$}}^\mu ~.
\end{eqnarray}
The 
scalar self interaction term \cite{schaffnerprc,glendenning,boguta} is, 
\begin{equation}
U(\sigma)~=~\frac13~g_1~m_N~(g_{\sigma N}\sigma)^3~+~ \frac14~g_2~
(g_{\sigma N}\sigma)^4~,
\end{equation}

and

\begin{equation}
\omega_{\mu\nu}~=~\partial_\nu\omega_\mu~-~\partial_\mu\omega_\nu~,
\end{equation}
\begin{equation}
\boldsymbol\rho_{\mu\nu}~=~\partial_\nu\boldsymbol\rho_\mu~-
~\partial_\mu\boldsymbol\rho_\nu~.
\end{equation}
In mean field approximation, the effective mass of baryons $B$ is
\begin{equation}
m_B^*~=~m_B~-~g_{\sigma B} \sigma~,
\end{equation}
where $\sigma$ is given by its ground state expectation value
\begin{equation}
\sigma~=~\frac1{m_\sigma^2}\left(\sum_B g_{\sigma B} ~ n_S^B
~-~\frac{\partial U}{\partial \sigma}\right)~.
\end{equation}
The scalar density is given by
\begin{equation}
n_S^B~=~\frac2{(2\pi)^3} \int_0^{k_{F_B}} \frac{m_B^*}{\sqrt{k_B^2+m_B^{*^2}}} 
d^3k_B.
\end{equation}
The chemical potential for baryons $B$ is
\begin{equation}
\mu_B~=~\sqrt{k_{F_B}^2+m_B^{*^2}}~+~ \omega^0~g_{\omega B}
~+~\rho_3^0~ g_{\rho B}~I_{3B},
\end{equation}
where $I_{3B}$ is the isospin projection and
\begin{equation}
\omega^0~=~\frac1{m_\omega^2}\sum_B g_{\omega B}~ n_B,
\end{equation}
\begin{equation}
\rho_3^0~=~\frac1{m_\rho^2}\sum_B g_{\rho B} ~I_{3B}~n_B.
\end{equation}
The total baryon number density is $n_b = \sum_B n_B$.

Now we consider the effects of strong magnetic 
fields on hyperon matter. The motion of charged particles in a magnetic field 
is Landau quantized in the plane perpendicular to the direction of the field.
We solve Dirac equations for charged particles using the gauge corresponding to
the constant magnetic field $B_m$ along the
$z$ axis as $A_0 = 0$, $\vec A = (0,xB_m,0)$. In the presence of a constant 
magnetic field, the Lagrangian density for protons is taken from 
Ref.\cite{cbsI}. The positive energy solutions for protons are
\begin{equation}
\psi_\alpha= \frac{\left(\frac{\sqrt{b}}{2^\nu \nu!\sqrt{\pi}}\right)^{1/2}}
{\sqrt{L_y L_z}} e^{-\xi^2/2}~e^{-i(\epsilon t~-~k_y y~-~k_z z)}~{\cal U}_{\alpha,\nu}
(k,x),
\end{equation}
with
$\xi~=~\sqrt{b}\left(x-\frac{k_y}{qB_m}\right)$ and $b=qB_m$.

The positive energy spinors, ${\cal U}_{\nu}(k,x)$, \cite{bcdp,Yak,Lei,goyal}
are given by 

\begin{equation}
{\cal U}_{\uparrow,\nu}(k,x)~=\sqrt{\epsilon^{'} + m^*_p}
~\left(
       \begin{array}{c}
       H_\nu(\xi) \\
       0 \\
       \frac{p_z}{\epsilon^{'} +m^*_p} H_\nu(\xi) \\
       \frac{-\sqrt{2\nu b}}{\epsilon^{'} + m^*_p} H_{\nu+1}(\xi)~\\
       \end{array}
\right), 
\label{psimagp}
\end{equation}
and
\begin{equation}
{\cal U}_{\downarrow,\nu}(k,x)~=\sqrt{\epsilon^{'} + m^*_p}
~\left(
       \begin{array}{c}
       0 \\
       H_\nu(\xi) \\
       \frac{-\sqrt{2\nu b}}{\epsilon^{'} + m^*_p} H_{\nu-1}(\xi) \\
       \frac{-p_z}{\epsilon^{'} + m^*_p} H_\nu(\xi)~\\
       \end{array}
\right)~,
\label{psimagm}
\end{equation}
where $\epsilon^{'} = \sqrt{p_z^2 + m_p^{*2} + 2 \nu q B_m}$.

The proton number density $n_p$ and scalar density $n_S^p$ are given by 
\cite{cbsI}

\begin{equation}
n_p~=~\frac{qB_m}{2\pi^2} \sum_{\nu=0}^{\nu_{max}} 
g_\nu k_p(\nu),
\end{equation}

\begin{equation}
n_S^p~=~\frac{qB_m}{2\pi^2}   
m_p^*\sum_{\nu=0}^{\nu_{max}} g_{\nu} 
\ln \frac{k_p(\nu)+\mu_p^*}{\sqrt{(m_p^{*^2} + 2\nu qB_m)}}~,
\end{equation}
where $\mu_B^*~=~\sqrt{k_{FB}^2+m_B^{*^2}}$ and
$k_p(\nu)~=~\sqrt{k_{F_p}^2 - 2\nu qB_m}$.
Maximum number of Landau levels populated is denoted by $\nu_{max}$ and
the Landau level degeneracy  $g_\nu$ is 1 for $\nu=0$ and 2 for $\nu>0$.
Similarly, we treat noninteracting electrons and muons in constant magnetic 
fields.

The total energy density of neutron star matter is 
\begin{eqnarray}
\varepsilon &=&  \frac12 m_\sigma^2 \sigma^2 + U(\sigma)
 + \frac12 m_\omega^2 \omega^{0^2}~+~\frac12 m_\rho^2 \rho_3^{0^2}
 + \sum_{B=n,\Lambda} 
\frac1{8 \pi^2} \left(2 {k_{F_B}\mu_B^{*^3}}
 - k_{F_B}m_B^{*^2}\mu_B^*
 - m_B^{*^4}~\ln \frac{k_{F_B}+\mu_B^*}{m^*_B} \right) \nonumber\\
&& +  \frac{qB_m}{(2\pi)^2} \sum_{\nu=0}^{\nu_{max}} g_{\nu} 
\left(k_p(\nu)\mu_p^*~+~( m_p^{*^2} + 2\nu qB_m) 
\ln \frac{k_p(\nu)+\mu_p^*}{\sqrt {(m_p^{*^2} + 2\nu qB_m)}}\right) \nonumber\\
&& + \frac{qB_m}{(2\pi)^2} \sum_{l=e,\mu} \sum_{\nu=0}^{\nu_{max}} 
\left(k_l(\nu)\mu_l~+~(m_l^2 + 2\nu qB_m) 
\ln \frac{k_l(\nu)+\mu_l}{\sqrt{(m_l^2 + 2\nu qB_m)}}\right)
+ \frac {B_m^2}{8\pi}~.
\end{eqnarray}

Similarly the total pressure of the system is given by
\begin{eqnarray}
P = - \frac12 m_\sigma^2 \sigma^2 - U(\sigma)
 + \frac12 m_\omega^2 \omega^{0^2}~+~\frac12 m_\rho^2 \rho_3^{0^2}
+ \frac{1}{3}\sum_{B=n,\Lambda} \frac{2J_B+1}{2\pi^2}
\int_0^{k_{F_B}} \frac{k^4 \ dk}{(k^2+{m_B^{*}}^2)^{1/2}} \nonumber\\
+ \frac{qB_m}{(2\pi)^2} \sum_{\nu=0}^{\nu_{max}} 
\left\{k_p(\nu)\mu_p^*~-~(m_p^{*^2} + 2\nu qB_m) 
\ln \frac{k_p(\nu)+\mu_p^*}{\sqrt{(m_p^{*^2} + 2\nu qB_m)}}\right\} \nonumber\\
+ \frac{qB_m}{(2\pi)^2} \sum_{l=e,\mu} \sum_{\nu=0}^{\nu_{max}} 
\left\{k_l(\nu)\mu_l~-~(m_l^2 + 2\nu qB_m) 
\ln \frac{k_l(\nu)+\mu_l}{\sqrt{(m_l^2 + 2\nu qB_m)}}\right\}
+ \frac {B_m^2}{8\pi}~,
\end{eqnarray}
where $k_l(\nu)~=~\sqrt{k_{F_l}^2 - 2\nu qB_m}$~.
The relation between pressure and energy density defines the equation of state
(EoS).
\section{\label{bv}Bulk viscosity}
The macroscopic compression (or expansion) of a fluid element leads to 
departure from chemical equilibrium. Non-equilibrium processes cause 
dissipation of energy which is the origin of bulk viscosity in neutron stars. 
Weak interaction processes bring the system back to equilibrium. 
In this calculation, we consider the non-leptonic reaction
\begin{equation}
n~+~p~\longrightarrow~p~+~\Lambda~, 
\label{lreac}
\end{equation}
as well as direct Urca (dUrca) processes which are represented by 
\begin{equation}
n~\longrightarrow~p~+~l^-~+~\bar{\nu_l},
\label{urca}
\end{equation}
where $l$ stands for $e$ or $\mu$. When the chemical equilibrium is achieved,
chemical potentials involved in above reactions
satisfy $\mu_n-\mu_\Lambda=0$ and $\mu_n-\mu_p-\mu_l=0$ 
respectively. In this case the forward and reverse reaction rates, $\Gamma_f$
and $\Gamma_r$ are same. The departure from chemical equilibrium due to 
macroscopic perturbation gives rise to the difference between forward and 
reverse reaction rates, $\Gamma = \Gamma_f-\Gamma_r\, \neq\, 0$. For a rotating
neutron star, the r-mode oscillation provides the macroscopic perturbation 
which drives the system out of chemical equilibrium.

The real part of bulk viscosity coefficient can be written as \cite{lindblom}
\begin{equation}
\zeta~=~-\frac {n_b^2\tau}{1+(\omega \tau)^2} \left(\frac {\partial P}
{\partial n_n}\right) \frac {d{\bar x_n}}{dn_b},
\label{zetag}
\end{equation}
where ${\bar x_i}=n_i/n_b$ is the equilibrium fraction of $i$-th species, 
$\omega$ is the
angular velocity of $(l,m)$ r-mode and $\tau$ is 
the microscopic relaxation time. For a neutron
star rotating with angular velocity $\Omega$, the angular velocity ($\omega$) 
of $(l,m)$ r-mode is given by 
\begin{equation}
\omega~=~\frac {2m}{l(l+1)} \Omega.
\end{equation}
We are interested in $l=m=2$ r-mode in this calculation.
The relaxation time is given by
\begin{equation}
\frac 1{\tau}~=~\frac {\Gamma}{\delta \mu} \frac {\delta \mu}{n_b \delta x_n}
\label{reactau}
\end{equation}
where $\delta \mu$ refers to the chemical imbalance. Here
$\Gamma$ is the total reaction rate.

The partial derivative of pressure with respect to neutron number density
can be evaluated from the EoS under consideration as
\begin{equation}
\frac{\partial P}{\partial n_n}~=~\frac {k_{F_n}^2}{3 \mu^*_n} \, 
-~\frac{\frac{g_{\sigma N}}{m_\sigma}
\frac{m_n^*}{\mu_n^*}
}D \sum_B n_B \frac{g_{\sigma B}}{m_\sigma} \frac{m^*_B}{\mu^*_B} 
~+~g_{\omega N} \omega^0~+~g_{\rho N} I_{3n} \rho_3^0, 
\label{delpdelnn}
\end{equation}

\begin{equation}
D~=~1+\sum_B \left(\frac{g_{\sigma B}}{m_\sigma}\right)^2 \, 
\frac{\partial n_S^B} {\partial m^*_B}~+~\frac 1{m_\sigma^2} 
\frac{\partial^2 U}{\partial \sigma^2}.
\label{Denom}
\end{equation}
The total derivative $dx_n/dn_b$ can be evaluated numerically.

Now, we calculate relaxation times for above mentioned processes in presence of
magnetic field $B_m$ using the EoS as described in section \ref{eos}. 

\subsection{\label{hyperon}Non-leptonic process}

Here we consider the non-leptonic process given by
Eq. (\ref{lreac}). In this case, only protons are affected by magnetic
fields. The reaction rate is given by 

\begin{eqnarray}
\Gamma~=~\int \frac{V d^3k_n}{(2\pi)^3} \int \frac{L_z dk_{p_{iz}}}{2\pi}
\int_{-\frac{bL_x}{2}}^{\frac{bL_x}{2}} \frac{L_y dk_{p_{iy}}}{2\pi} \int 
\frac{L_z dk_{p_{fz}}}{2\pi} \int_{-\frac{bL_x}{2}}^{\frac{bL_x}{2}} 
\frac{L_y dk_{p_{fy}}}{2\pi} \int \frac{V d^3k_\Lambda}{(2\pi)^3}~W_{fi}
\nonumber \\ 
~\times F(\epsilon_n,\epsilon_{p_i},\epsilon_{p_f},\epsilon_\Lambda),
\label{gammal1}
\end{eqnarray} 
$k_{p_{iz}}$ and $k_{p_{fz}}$ being the $z$ component of momenta of initial 
and final protons respectively and $k_n$ and $k_{\Lambda}$ denote momenta of
neutrons and $\Lambda$ hyperons. The Pauli blocking factor is given by
\begin{equation}
F(\epsilon_n,\epsilon_{p_i},\epsilon_{p_f},\epsilon_\Lambda) =
f(\epsilon_n) f(\epsilon_{p_i}) \{1-f(\epsilon_{p_f})\} 
\{1-f(\epsilon_\Lambda)\}~
- f(\epsilon_\Lambda) f(\epsilon_{p_f})
\{1-f(\epsilon_{p_i})\} \{1-f(\epsilon_n)\}~,
\end{equation}
with the Fermi distribution function at temperature $T$
\begin{equation}
f(\epsilon_i)~=~\frac 1{1~+~e^{\frac{\epsilon_i~-~\mu}{kT}}}.
\end{equation}
The matrix element $W_{fi}$ is given by 
\begin{equation}
W_{fi}~=~\frac 1{V^3 (L_y L_z)} \frac{(2\pi)^3}{16 \epsilon_n \epsilon_{p_i}
\epsilon_{p_f} \epsilon_\Lambda} ~|{\cal M}|^2 ~e^{-Q^2}~ \delta(\epsilon) 
\delta(k_y) \delta(k_z)~, 
\label{matel}
\end{equation}
where
\begin{equation}
Q^2~=~\frac{(k_{nx}-k_{\Lambda x})^2~+~(k_{p_{iy}}-k_{p_{fy}})^2}{2b} ~~~~~~~~
{\rm and}~~~~~~~~\delta(k)~\equiv~\delta(k_n+k_{p_i}-k_{p_f}-k_\Lambda).
\end{equation}
The invariant amplitude squared for the process is
\begin{eqnarray}
|{\cal M}|^2~=~4 G_F^2 \sin^2 2\theta_c \left[2 m^*_n {m^*_p}^2 m^*_\Lambda
(1-g_{np}^2) (1-g_{p\Lambda}^2) \right. \nonumber \\ \left.  -~m^*_n m^*_p (k_{p
_i}\cdot k_\Lambda)
(1-g_{np}^2) (1+g_{p\Lambda}^2)
~ - m^*_p m^*_\Lambda (k_n\cdot k_{p_f}) (1+g_{np}^2) (1-g_{p\Lambda}^2)
\right. \nonumber \\ \left. +~(k_n\cdot k_{p_i}) (k_{p_f}\cdot k_\Lambda) \{(1+g
_{np}^2)
(1+g_{p\Lambda}^2)~+~4 g_{np} g_{p\Lambda} \} \right. \nonumber \\
\left. + (k_n\cdot k_\Lambda) (k_{p_i}\cdot k_{p_f}) \{(1+g_{np}^2)
(1+g_{p\Lambda}^2)~-~4 g_{np} g_{p\Lambda} \}\right].
\end{eqnarray}

In calculating the matrix element given by Eq. (\ref{matel}) we use the
solutions of Dirac equation for protons in magnetic field given by 
Eqs. (\ref{psimagp}) and (\ref{psimagm}). We also assume that the magnetic 
field is so strong that only zeroth Landau level is populated.
Now we integrate
over $k_{p_{iy}}$ and $k_{p_{fy}}$ using $\delta(k_y)$ and obtain
\begin{eqnarray}
\Gamma~=~\frac{L_yL_z}{(2\pi)^7 V 16}~b L_x~\int d^3k_n \int dk_{p_{iz}}
\int dk_{p_{fz}} \int d^3k_\Lambda \left(\frac {|{\cal M}|^2}{\epsilon_n 
\epsilon_{p_i} \epsilon_{p_f} \epsilon_\Lambda}\right)_{\delta(k_y)}
\nonumber \\
\times e^{-[(k_{nx}-k_{\Lambda x})^2+(k_{ny}-k_{\Lambda y})^2]/2b}
~F(\epsilon_n,\epsilon_{p_i},\epsilon_{p_f},\epsilon_\Lambda)
~\delta(\epsilon_n+\epsilon_{p_i}-\epsilon_{p_f}-\epsilon_\Lambda) \delta(k_z).
\end{eqnarray}
Here the subscript $\delta(k_y)$ denotes that this condition has been imposed
on the quantity within the parenthesis. Next we perform the integration over 
${\boldsymbol k}_n$ and ${\boldsymbol k}_\Lambda$ and write $d^3 k~=~k^2
~dk~d(\cos \theta)~d\phi$. The delta function of z-components 
of momenta implies $k_{nz}+k_{p_{iz}}=k_{p_{fz}}+k_{\Lambda z}$. Here we note 
that when protons occupy only the zeroth Landau level, they have momenta 
along the direction of magnetic field {\it i.e.} in $z$ direction. 
Hence we have $k_{pz}=k_{F_p}$.
Then depending upon whether the initial and final protons are moving in 
the same or opposite direction we have $k_{\Lambda z}-k_{nz}=0 $ or 
$k_{\Lambda z}-k_{nz}=2k_{F_p}$. Next we perform the angle integrations 
using $\delta(k_z)$ and change variable $k$ to $\epsilon$ to get
\begin{eqnarray}
\Gamma~=~\frac b{(2\pi)^5 8} \int d\epsilon_n d\epsilon_{pi}
d\epsilon_{pf}
d\epsilon_\Lambda \frac{k_{F_\Lambda}}{k_{F_p}k_{F_p}}
\left((|{\cal M}|^2)_{\theta_{int}}
\right)_{\delta(k_y),\delta(k_z)} \nonumber \\ \times \left[\Theta\{(k_{F_n}-k_{F_\Lambda})^2\}
e^{-[(k_{F_n}-k_{F_\Lambda})^2]/2b}
~ +\Theta\{(k_{F_n}-k_{F_\Lambda})^2-4k_{F_p}^2\}
e^{-[(k_{F_n}-k_{F_\Lambda})^2-4k_{F_p}^2]/2b}\right]\nonumber\\
\times
F(\epsilon_n,\epsilon_{p_i} ,\epsilon_{p_f},\epsilon_\Lambda)
\delta(\epsilon_n+\epsilon_{p_i}-\epsilon_{p_f}-\epsilon_\Lambda).
\label{gammal4}
\end{eqnarray}
Here the subscript $\theta_{int}$ denotes the angle integrated value. As
particles reside near their Fermi surfaces in a degenerate matter 
we replace momenta and energies under 
integration by their respective values at their Fermi surfaces. 

The matrix element squared is rewritten as,
\begin{eqnarray}
\left((|{\cal M}|^2)_{\theta_{int}}\right)_{\delta(k_y),\delta(k_z)} 
~=~4 G_F^2 \sin^2 2\theta_c \left[2 m^*_n {m^*_p}^2 m^*_\Lambda (1-g_{np}^2)
(1-g_{p\Lambda}^2) \nonumber \right.\\
\left.~-~m^*_n m^*_p \mu_p \mu_\Lambda (1-g_{np}^2) 
(1+g_{p\Lambda}^2) -~m^*_p m^*_\Lambda \mu_n \mu_p (1+g_{np}^2) 
(1-g_{p\Lambda}^2)~\nonumber \right. \\
\left. + ~\mu_n \mu_p^2 \mu_\Lambda \{(1+g_{np}^2) (1+g_{p\Lambda}^2)~+
~4 g_{np} g_{p\Lambda}\} \nonumber \right. \\
\left. +~ \mu_n \mu_p^2 \mu_\Lambda 
\left(1-\frac{k_{F_p}^2}{\mu_p^2}\right)
\{(1+g_{np}^2) (1+g_{p\Lambda}^2)~-
~4 g_{np} g_{p\Lambda}\right].
\end{eqnarray}
As $\delta \mu << kT$, the energy integration  of Eq. 
(\ref{gammal4}) can be written as \cite{lindblom}
\begin{equation}
\int d\epsilon_n d\epsilon_{p_i} d\epsilon_{p_f} d\epsilon_\Lambda
~F(\epsilon_n,
\epsilon_{p_i},\epsilon_{p_f},\epsilon_\Lambda) ~\delta(\epsilon_n
+\epsilon_{p_i}-\epsilon_{p_f}-\epsilon_\Lambda)~=~(kT)^2~\frac{2\pi^2}3
~\delta \mu.
\end{equation}
Finally we get
\begin{eqnarray}
\Gamma~=~\frac 1{384 \pi^3}~\frac{qB_m k_{F_\Lambda}}{k_{F_p}^2}~
\left( (|{\cal M}|^2)_{\theta_{int}}\right)_{\delta(k_y),\delta(k_z)} 
\left[\Theta\{(k_{F_n}-k_{F_\Lambda})^2\}
e^{-[(k_{F_n}-k_{F_\Lambda})^2]/2b} \right. \nonumber \\~
\left. +\Theta\{(k_{F_n}-k_{F_\Lambda})^2-4k_{F_p}^2\}
e^{-[(k_{F_n}-k_{F_\Lambda})^2-4k_{F_p}^2]/2b}\right]
(kT)^2~\delta \mu.
\label{gammaf}
\end{eqnarray}

The expression of the reaction rate for a zero magnetic field is given 
by\cite{lindblom}
\begin{equation}
\Gamma~=~\frac 1{192 \pi^3}~\langle |{\cal M}|^2\rangle~k_{F_\Lambda}~
(kT)^2~\delta \mu,
\label{gammafwm}
\end{equation}
where the angle averaged matrix element squared is same as given by 
\cite{lindblom}.

Now the quantity $\delta \mu/\delta x_n$ in Eq. (\ref{reactau}) is to be 
evaluated under the condition of total baryon number conservation 
\cite{lindblom}
\begin{equation}
\delta n_n~+~\delta n_\Lambda~=~0,
\end{equation}
which leads to
\begin{equation}
\frac{\delta \mu}{\delta x_n}~=~\alpha_{nn}~-~\alpha_{n\Lambda}~-
~\alpha_{\Lambda n}~+~\alpha_{\Lambda \Lambda},~~~~{\rm with}~~~~ 
\alpha_{ij}~=~\frac{\partial \mu_i}{\partial n_j}.
\label{delmudelxn}
\end{equation}
Further we have
\begin{equation}
\alpha_{ij}~=~\frac{\pi^2}{k_{F_i} \mu^*_i} 
\,\delta_{ij}~-~\frac{m^*_i}{\mu^*_i} \, \frac{\left(\frac{g_{\sigma i}}
{m_\sigma}\right) \left(\frac{g_{\sigma j}}{m_\sigma}\right) 
\frac{m_j^*}{\mu_j^*}}D ~+~ \frac1 {m_\omega^2} g_{\omega i} g_{\omega j}
~+~\frac1{m_\rho^2} \; g_{\rho i} I_{3 i} \; g_{\rho j} I_{3 j}.
\label{alphagen} 
\end{equation}
Here $D$ is the same as given by Eq. (\ref{Denom}).
Next we evaluate the relaxation time of the non-leptonic reaction 
at a given baryon density using Eq. (\ref{reactau}) along with 
Eqs. (\ref{gammaf}), (\ref{delmudelxn}) and (\ref{alphagen}). 

As soon as we know the relaxation time, 
we can calculate the bulk viscosity coefficient $\zeta$ due to the non-leptonic 
interaction at a given baryon density from Eq. (\ref{zetag}). 

\subsection{\label{URCA}Leptonic processes}
Here we consider dUrca processes involving nucleons, electrons and muons in
a magnetic field. The forward reaction rate for this process is then given 
by \cite{bcdp,Yak,Lei}
\begin{eqnarray}
\Gamma_f &=& \int \frac{V d^3k_n}{(2\pi)^3}  \int \frac{V d^3k_\nu}{(2\pi)^3}
\int \frac{L_z dk_{zp}}{2\pi}
\int_{-\frac{bL_x}{2}}^{\frac{bL_x}{2}} \frac{L_y dk_{yp}}{2\pi} \int 
\frac{L_z dk_{zl}}{2\pi} \int_{-\frac{bL_x}{2}}^{\frac{bL_x}{2}} 
\frac{L_y dk_{yl}}{2\pi}~W_{fi} \nonumber\\
&&\times F(\epsilon_n,\epsilon_p,\epsilon_l).
\label{gammau1}
\end{eqnarray} 
Here $F(\epsilon_n,\epsilon_p,\epsilon_l)$ is given by
\begin{equation}
F(\epsilon_n,\epsilon_p,\epsilon_l)~=
~f(\epsilon_n) \{1-f(\epsilon_p)\} \{1-f(\epsilon_l)\}.
\label{fermi}
\end{equation}
Using the solutions of Dirac equations for protons and electrons in magnetic 
field, we obtain the matrix element 
\begin{equation}
W_{fi}~=~\frac {(2\pi)^3}{V^3 (L_y L_z)} |{\cal M}|^2 \delta(\epsilon)
\delta(k_y) \delta(k_z)~.
\end{equation}

Firstly we treat the case following the prescription of Baiko and Yakovlev 
\cite{Yak} when protons and electrons populate large numbers of Landau levels.
In this case, we have 
\begin{equation}
\sum_{s_n,s_p} |{\cal M}|^2 = 2 G_F^2 \cos^2 \theta_c (1 + 3G_A^2) F^2~,
\end{equation}
where $F$ is Laguerre functions for both protons and electrons \cite{Yak}.
The forward reaction rate is given by,
\begin{equation}
\Gamma_f = \frac{32 \pi G_F^2 \cos^2 \theta_c m_n^* m_p^* \mu_l} {(2\pi)^5} 
R_B^{qc} \int d\epsilon_\nu \epsilon_\nu^2 J(\epsilon_\nu)~,
\end{equation}
where 
\begin{eqnarray}
R_B^{qc} &=& 2 \int \int_{-1}^{1} d\cos \theta_p d\cos \theta_l 
\frac{{K_{F_p}} {K_{F_l}}} {4b} F_{{N_p},{N_l}}^2 (u)
\Theta({k_{F_n}}-|{k_{F_p} \cos \theta_p + k_{F_l} \cos \theta_l}|)~,
\end{eqnarray}
and 
\begin{eqnarray}
J(\epsilon_\nu)&=&\int d\epsilon_n d\epsilon_p d\epsilon_l
F(\epsilon_n,\epsilon_p,\epsilon_l)
\delta(\epsilon_n-\epsilon_p-\epsilon_l-\epsilon_\nu)~, \nonumber\\
&&=\frac {(k T)^2}2\:
\frac{\pi^2+(\epsilon_\nu/k T)^2}{1+e^{\epsilon_\nu/k T}}.
\end{eqnarray}
As there is chemical imbalance due to the perturbation, the reverse reaction
rate ($\Gamma_r$) differs from the forward reaction rate and the net reaction
rate is given by \cite{Han,Yak}
\begin{eqnarray}
\Gamma_l &=&
\frac{32 \pi G_F^2 \cos^2 \theta_c m_n^* m_p^* \mu_l} {(2\pi)^5} 
R_B^{qc} \int d\epsilon_\nu \epsilon_\nu^2
\{J(\epsilon_\nu-\delta \mu)-J(\epsilon_\nu+\delta \mu)\}.
\label{deltagwm}
\end{eqnarray}

One important aspect of dUrca process is the opening of this channel in the
forbidden regime $K_{F_n} > K_{F_p} + K_{F_l}$ which was otherwise closed in 
field free case \cite{Yak}. The dUrca process also operates in the allowed 
domain $K_{F_p} + K_{F_l} > K_{F_n}$ in the presence of a magnetic field. We 
adopt fitting formulas for $R_B^{qc}$ in both domains as given by 
Ref.\cite{Yak}.

Next we focus on the case when both protons and electrons populate 
zeroth Landau levels \cite{bcdp,Yak,Lei}. In this case we write the matrix 
element as
\begin{equation}
W_{fi}~=~\frac {(2\pi)^3}{V^3 (L_y L_z)} \frac1{16 \epsilon_n \epsilon_\nu
\epsilon_p \epsilon_e} ~|{\cal M}|^2 ~e^{-Q^2}~ \delta(\epsilon) 
\delta(k_y) \delta(k_z)~, 
\label{matelu}
\end{equation}
\begin{equation}
Q^2~=~\frac{(k_{nx}-k_{\nu x})^2~+~(k_{p y}+k_{l y})^2}{2b}.
\end{equation}
In a magnetic field neutrons will be 
polarized because of their anomalous magnetic moments. Hence for two 
different spin states of neutrons, matrix elements should be 
evaluated separately. The invariant amplitude squared is then 
$|{\cal M}|^2= |{\cal M_+}|^2+|{\cal M_-}|^2$, where
\begin{eqnarray}
|{\cal M_\pm }|^2=\frac{G_F^2}2\sum_s\{\bar {\cal V}_{\nu s} (k_\nu )
(1+\gamma^5)
\gamma_\nu {\cal U}_{l-}(k_l)\}\{\bar{\cal U}_{n\pm}(k_n)(1-g_{np}\gamma^5)
\gamma^\nu {\cal U}_{p+}(k_p)\} \\
\times \{\bar{\cal U}_{p+}(k_p)\gamma^\mu(1+g_{np}\gamma^5){\cal U}_{n\pm}(k_n)\}
\{\bar{\cal U}_{l-}(k_e)\gamma_\mu(1-\gamma^5){\cal V}_{\nu s}(k_\nu)\}~,
\end{eqnarray}
and $\pm$ signs denote the up and down spins respectively.
The spinors for non-relativistic neutrons are given by 
\begin{equation}
{\cal U}_{n\pm}=\sqrt{\epsilon_n+m_n^*}\left(\begin{array}{c}
				\chi_\pm  \\
				0 
				\end{array}\right),
\label{spinucl}
\end{equation}
where 

\begin{equation}
\chi_+=\left(\begin{array}{c}
		1 \\
		0
		\end{array}\right)~~~~~~{\rm and}~~~~~
\chi_-=\left(\begin{array}{c}
		0 \\
		1
		\end{array}\right).
\end{equation}
For non-relativistic protons in the zeroth Landau level, the spinor
has the same form as given by Eq. (\ref{spinucl}).
For spin down relativistic leptons in the zeroth Landau level, the spinor 
is given by
\begin{equation}
{\cal U}_{l-}=\sqrt{\epsilon_l+m_l}\left(\begin{array}{c}
					0 \\
					1 \\
					0 \\
					\frac{-p_{lz}}{\epsilon_l+m_l}
					\end{array}\right)
\end{equation}
For spin up and down neutrons, invariant amplitudes squared are
\begin{equation}
|{\cal M}_+|^2~=~8 G_F^2 \cos^2 \theta_c m_n^* m_p^* (1+g_{np})^2
(\epsilon_l+p_l)(\epsilon_\nu+p_{\nu z})~,
\end{equation}
and
\begin{equation}
|{\cal M}_-|^2~=~32 G_F^2 \cos^2 \theta_c m_n^* m_p^* g_{np}^2
(\epsilon_l+p_l)(\epsilon_\nu-p_{\nu z})~.
\end{equation}

Following the same procedure as described in subsection \ref{hyperon} 
and neglecting the neutrino momenta in momentum conserving delta functions,
the final expression of forward reaction rate $\Gamma_f$ is given by 
\begin{eqnarray}
\Gamma_f&=&\frac b{(2\pi)^5 8} \frac{m_n^* m_p^* \mu_l}{k_{F_p} k_{F_l}} 
\left[\left(|{\cal M_+}|^2_d\right)_{\delta(k_y), \delta(k_z)} + 
\left(|{\cal M_-}|^2_d\right)_{\delta(k_y), \delta(k_z)}\right] \nonumber\\ 
&&\times \left[\Theta\{k_{F_n}^2-(k_{F_p}-k_{F_l})^2\}
e^{-[k_{F_n}^2-(k_{F_p}-k_{F_l})^2]/2b} \right. 
\left. +\Theta\{k_{F_n}^2-(k_{F_p}+k_{F_l})^2\}
e^{-[k_{F_n}^2-(k_{F_p}+k_{F_l})^2]/2b} \right] \nonumber \\
&&\times \int d\epsilon_\nu \epsilon_\nu^2 \int d\epsilon_n 
d\epsilon_p d\epsilon_l F(\epsilon_n,\epsilon_p,\epsilon_l) 
\delta(\epsilon_n-\epsilon_p-\epsilon_l-\epsilon_\nu)~,  
\end{eqnarray}
where 
\begin{eqnarray}
\left(|{\cal M}_+|^2_d\right)
_{\delta(k_y), \delta(k_z)}=8 G_F^2 \cos^2 \theta_c (1+g_{np})^2
\left(1+\frac{p_l}{\epsilon_l}\right)\left(1+\frac{p_{\nu z}}{\epsilon_\nu}\right)~.
\end{eqnarray}
Similarly we have,
\begin{eqnarray}
\left(|{\cal M}_-|^2_d\right)
_{\delta(k_y), \delta(k_z)}=32 G_F^2 \cos^2 \theta_c {g_{np}^2}
\left(1+\frac{p_l}{\epsilon_l}\right)\left(1-\frac{p_{\nu z}}{\epsilon_\nu}\right)~.
\end{eqnarray}
It is to be noted that z-component of neutrino momentum is smaller than its 
energy. We obtain
\begin{eqnarray}
\Gamma_f &=& \frac b{(2\pi)^5 8} \frac{m_n^* m_p^* \mu_l}{k_{F_p} k_{F_l}} 
\left[
\left(|{\cal M_+}|^2_d\right) _{\delta(k_y), \delta(k_z)} + 
\left(|{\cal M_-}|^2_d\right) _{\delta(k_y), \delta(k_z)}\right] \nonumber\\ 
&&\times \left[ \Theta\{k_{F_n}^2-(k_{F_p}-k_{F_l})^2\}e^{-[k_{F_n}^2-(k_{F_p}
-k_{F_l})^2]/2b} \right. 
\left. +
\Theta\{k_{F_n}^2-(k_{F_p}+k_{F_l})^2\}e^{-[k_{F_n}^2-(k_{F_p}+k_{F_l})^2]/2b} \right] \nonumber\\
&&\times \int d\epsilon_\nu \epsilon_\nu^2 
J(\epsilon_\nu). 
\end{eqnarray}
Now if the reverse reaction rate is $\Gamma_r$ and 
there is slight departure from chemical equilibrium $\delta \mu$, then the 
net reaction rate is \cite{Han}, 
\begin{eqnarray}
\Gamma_l&=&\Gamma_r-\Gamma_f=
\frac b{(2\pi)^5 8} \frac{m_n^* m_p^* \mu_l}{k_{F_p} k_{F_l}} 
\left[
\left(|{\cal M_+}|^2_d\right) _{\delta(k_y), \delta(k_z)} + 
\left(|{\cal M_-}|^2_d\right) _{\delta(k_y), \delta(k_z)}\right] \nonumber\\ 
&&\times \left[\Theta\{k_{F_n}^2-(k_{F_p}-k_{F_l})^2\}e^{-[k_{F_n}^2-(k_{F_p}-
k_{F_l})^2]/2b} \right. 
\left. +\Theta\{k_{F_n}^2-(k_{F_p}+k_{F_l})^2\}e^{-[k_{F_n}^2-(k_{F_p}
+k_{F_l})^2]/2b} \right] \nonumber\\
&&\times \int d\epsilon_\nu \epsilon_\nu^2 
\{J(\epsilon_\nu-\delta \mu)-J(\epsilon_\nu+\delta \mu)\}. 
\end{eqnarray}
Using the following result from Ref. \cite{Han}
\begin{equation}
\int d\epsilon_\nu \epsilon_\nu^2 
\{J(\epsilon_\nu-\delta \mu)-J(\epsilon_\nu+\delta \mu)\}=\frac{17 (\pi k T)^4}
{60} \, \delta \mu,  
\end{equation}
we get
\begin{eqnarray}
\Gamma_l &=&\frac {17 qB_m}{480 \pi}\frac{m_n^* m_p^* \mu_l}{k_{F_p} k_{F_l}}
G_F^2 \cos^2 \theta_c \left(1+\frac{p_l}{\epsilon_l}\right)
\left[\frac{1}{4} (1+g_{np})^2 + g_{np}^2 \right] \nonumber\\ 
&&\times \left[\Theta\{k_{F_n}^2-(k_{F_p}-k_{F_l})^2\}e^{-[k_{F_n}^2-(k_{F_p}-k_{F_l})^2]/2b} \right. 
\left. +\Theta\{k_{F_n}^2-(k_{F_p}+k_{F_l})^2\}e^{-[k_{F_n}^2-(k_{F_p}+k_{F_l})^2]/2b} \right] \nonumber \\
&&\times (k T)^4 \, \delta \mu.
\label{deltags}
\end{eqnarray}

The zero magnetic field result is given by
\begin{equation}
\Gamma_l (B_m=0) = \frac{17}{240 \pi} m_n^* m_p^* \mu_l 
(|{\cal M}|^2_d)_{\theta_{int}}
(k T)^4 \delta \mu,
\end{equation}
where 
\begin{equation}
(|{\cal M}|^2_d)_{\theta_{int}}
=
G_F^2 \cos^2 \theta_c \left\{(1+g_{np})^2 \left(1-\frac{k_{F_n}}{m_n^*}\right)
+(1-g_{np})^2\left(1-\frac{k_{F_p}}{m_p^*}\right)
-(1-g_{np}^2) 
\right\}.
\end{equation}

\section{\label{rd}Results and discussion}
Nucleon-meson coupling constants of the model are obtained by 
reproducing the properties of nuclear matter such as binding energy 
$E/B=-16.3~MeV$, 
saturation density $n_0=0.153~fm^{-3}$, asymmetry energy coefficient 
$a_{asy}=32.5~MeV$ and incompressibility $K=240~MeV$ and
taken from Ref \cite{glend}. The coupling strength of $\Lambda$ hyperons with
$\omega$ mesons is determined from SU(6) symmetry of the quark model
\cite{sm,dg,sdgmgs}. The coupling strength of $\Lambda$ hyperons to $\sigma$ 
mesons is determined from the potential depth of $\Lambda$ hyperons in normal 
nuclear matter
\begin{equation}
U_\Lambda=-g_{\sigma \Lambda} \sigma+g_{\omega \Lambda} \omega_0.
\end{equation}
We take the potential depth $U_\Lambda=-30~MeV$ 
as obtained from the analysis of $\Lambda$ hypernuclei \cite{dg,fet}.

We adopt a profile of magnetic field given by \cite{bcpII},
\begin{equation}
B\left(n_b/n_0\right)=B_s+B_c\left(1-e^{-\beta \left(
\frac {n_b}{n_0} \right)^\gamma}\right).
\end{equation} 
We consider different values for central field $B_c = 10^{16}$ and 
$10^{17}$ G whereas surface field strength is 
taken as $B_s = 10^{14}$ G in this calculation. We chose
$\beta=0.01$ and $\gamma=3$. The magnetic field strength depends on 
baryon density in the above parameterization. Further the magnetic field at 
each density point is constant and uniform. The effects of anomalous magnetic 
moments of nucleons and contributions of the magnetic field to energy density 
and 
pressure are negligible because magnetic fields considered in this calculation 
are not too strong.

Numbers of Landau levels populated by electrons and protons, are
sensitive to the magnetic field strength and baryon density. 
As the field strength increases, the population of Landau levels decreases.
In a weak magnetic field, when many Landau levels are populated, 
we treat charged particles unaffected by the magnetic field. Further the 
effects of magnetic fields are most pronounced when only zeroth Landau 
levels are populated. 
Protons, electrons and muons populate zeroth Landau levels if 
central field strength $B_c \sim 10^{19}$ G. 
Figure 1 shows fractions of various particle species with 
normalised baryon density. We find large numbers of Landau levels of charged
particles even when the magnetic field reaches its central value $10^{17}$ G. 
Populations of charged particles are enhanced in the magnetic field due to
Landau quantization than those of field free case (not shown in the figure).
It is noted in Fig. 1 that the threshold density of $\Lambda$ hyperons is 
shifted to 1.7$n_0$ from its zero magnetic field value of 2.6$n_0$ 
because of phase space modifications of charged particles 
in a magnetic field. 

The variation of pressure with energy density in the presence of a 
magnetic field with central field strength $B_c = 10^{17}$ G (solid curve) 
is shown in Fig. 2. The dashed curve denotes the EoS without a magnetic field. 
The EoS in the presence of the magnetic
field becomes stiffer when charged particles are Landau quantised. 
Here magnetic field contributions to the energy density and pressure are 
insignificant.  

Now we compute the relaxation time for both non-leptonic and leptonic
reactions as given by Eq. (\ref{reactau}). To calculate the matrix element
we take $g_{np}=-1.27$ and $g_{p\Lambda}=-0.72$ \cite{lindblom}, 
and the Cabibbo angle ($\theta_c$) is given by $\sin \theta_c=0.222$. As we 
have already noted, charged particles populate many Landau levels 
in a magnetic field having central value $B_c = 10^{17}$ G 
over entire density range considered in our calculation. 
For the non-leptonic process, when protons populate large number of 
Landau levels, we use the field free expression of 
$\Gamma$ as given by Eq. (\ref{gammafwm}). For leptonic reactions we use the 
expression as given by Eq. (\ref{deltagwm}) when leptons and protons 
populate finite numbers of Landau levels. Chemical potentials and Fermi momenta
of constituent particles are obtained from the EoS. The partial derivative of 
chemical potentials with respect to baryon density can be calculated from the 
EoS. Using these inputs, we can compute relaxation times for both reactions 
as a function of baryon density at a particular temperature. Figure 3 displays
relaxation time ($\tau$) of the non-leptonic process involving $\Lambda$ 
hyperons in a magnetic field having its central value $B_c = 10^{17}$ G 
and at different temperatures as a function of normalised baryon 
number density. Here $\tau$ decreases with increasing 
baryon density. Further the relaxation time in a magnetic field increases with 
decreasing temperature as was earlier noted in the field free case 
\cite{Chat1}.

Relaxation times for dUrca reactions involving electrons and muons in a 
magnetic field with $B_c = 10^{17}$ G and at different temperatures are plotted
in Figs. 4 and 5 respectively. For leptonic processes, relaxation 
times are affected by the magnetic field. For the field free case, the dUrca 
process sets in at 1.4$n_0$. In the magnetic field, relaxation times due to 
dUrca reactions drop sharply from large values in the forbidden domain 
$K_{F_n} > K_{F_p} + K_{F_e}$. This is attributed to the behaviour of 
$R_B^{qc}$ which we discuss in details in connection with bulk viscosity due
to dUrca processes below. The forbidden domain joins with  
the allowed domain $K_{F_p}+K_{F_e} > K_{F_n}$ at a point from which relaxation
times increase with baryon density. Like the non-leptonic case, 
relaxation times for dUrca processes also increase with decreasing temperature.

Now we focus on the calculation of bulk viscosity due to the non-leptonic and 
leptonic processes. As soon as we know relaxation times of non-leptonic and 
leptonic reactions, we compute bulk viscosity coefficients for the respective 
processes from Eq. (\ref{zetag}). In this calculation we consider $l=m=2$ 
r-mode and hence $\omega=2/3\Omega$. Further we take $\Omega = 3000 s^{-1}$. In
the temperature regime considered here, we have always  $\omega\tau<<1$ for the
non-leptonic process involving $\Lambda$ hyperons. Therefore, we neglect that 
term in the denominator of Eq. (\ref{zetag}) to calculate the hyperon bulk 
viscosity. The partial derivative of pressure with respect to neutron number 
density is calculated from the EoS
using Eq. (\ref{delpdelnn}) and the total derivative of neutron fraction
with respect to baryon density is computed numerically from 
the EoS. As the relaxation time is a function of temperature, the bulk
viscosity coefficient $\zeta$ also depends on temperature. 
The bulk viscosity coefficient for the non-leptonic process in a magnetic field 
with $B_c = 10^{17}$ G (dashed curve) and in the absence of a magnetic field 
(solid curve) are exhibited as a function of normalised baryon number 
density in Fig. 6 at different temperatures. The non-leptonic reaction 
involves protons that populate many Landau levels in the
magnetic field with $B_c = 10^{17}$ G. In this case, we adopt the field free 
expression of the reaction rate as given by Eq. (\ref{gammafwm}) for the 
calculation of relaxation time and hyperon bulk viscosity coefficient in 
Eq. (\ref{zetag}). Therefore, the effects of magnetic field enter into hyperon 
bulk viscosity coefficient through the EoS which is modified by Landau
quantization of charged particles. In Fig. 6, we find hyperon bulk viscosity 
in the magnetic field is suppressed compared with the field free case.

We display bulk viscosity coefficient for the dUrca process in a magnetic field 
with $B_c = 10^{16}$ G and at a temperature $T = 10^{11}$ K as a function of
normalised baryon density in Fig. 7. In this case electrons and protons 
populate many Landau levels. The dotted line represents
the dUrca contribution in the forbidden domain $K_{F_n} > K_{F_p} + K_{F_e}$.
In this regime, reaction kinetics are characterised by two parameters 
$x = \frac{K_{F_n}^2 - {(K_{F_p} + K_{F_e})^2}}{K_{F_p}^2 {N_{F_p}^{-2/3}}}$
and $y = N_{F_p}^{2/3}$, where $N_{F_p}$ is the number of proton Landau levels. 
The dUrca reaction in the forbidden domain is an efficient process as long as
$x \leq 10$. This corresponds to baryon density $\leq$ 2.3$n_0$. The large 
enhancement of bulk viscosity coefficient in this domain is attributed to 
the behaviour of $R_B^{qc}$ \cite{Yak}. It was noted $R_B^{qc} = 1/3$ at $x=0$
and it becomes very small when $x>10$ \cite{Yak}. At $x=0$, the forbidden 
domain merges with the allowed domain $K_{F_p}+K_{F_e} > K_{F_n}$ of the dUrca 
process. The dUrca bulk viscosity in the allowed domain is shown
by the dash-dotted line. The result of zero field is shown by the solid 
line. The bulk coefficient increases with magnetic field in the allowed
domain at higher baryon densities.

Figure 8 and Figure 9 show bulk viscosity coefficients for dUrca 
processes involving electrons and  muons in the presence of the magnetic field 
with central value $B_c = 10^{17}$ G and at different temperatures as a 
function of normalised baryon density. In both cases 
contributions to bulk viscosity coefficients due to dUrca processes
come from the forbidden as well as allowed domains. As discussed above,
the forbidden domain merges with the allowed domain at $x=0$. 
For temperatures $T = 10^9$ and 
$10^{10}$ K, bulk viscosity coefficients due to dUrca processes increase 
with baryon density. However the bulk viscosity for $T = 10^{11}$ K initially 
decreases and later increases with baryon density. This behaviour can be 
understood in the following way. For dUrca processes at $10^{11}$ K, we have 
$\omega \tau < 1$. 
On the other hand, we find $\omega \tau > 1$ for dUrca processes at 
$10^9$ K and $10^{10}$ K. Consequently bulk viscosity 
coefficients have a $T^4$ dependence when $\omega \tau > 1$ whereas it has a 
$T^{-4}$ dependence when $\omega \tau < 1$. This inversion of temperature
dependence of dUrca bulk viscosity coefficients is not found in the case of
hyperon bulk viscosity. 

Finally, we point out what happens in case of superstrong fields. 
We find that charged particles populate zeroth Landau levels when
$B_c \sim 10^{19}$. Populations of charged particles are enhanced because of 
strong modification of their phase spaces. Further the EoS is modified due
to magnetic field contributions to the energy density and pressure. The strong 
magnetic field enhances the hyperon bulk viscosity compared with the field free
case. Similarly we note significant modification in bulk 
viscosity coefficients due to dUrca processes when leptons and protons are
in zeroth Landau levels. However, there is no observational evidence
for superstrong field $\sim 10^{19}$ G in neutron star's interior so far. 

\section{\label{sum}Summary}
We have investigated bulk viscosity of non-leptonic process involving $\Lambda$
hyperons and dUrca processes in the presence of strong magnetic fields. In
this calculation we consider magnetic fields with different central values 
$B_c = 10^{16}$ and $10^{17}$ G. The equation of state has been constructed 
using the relativistic field theoretical model. Many
Landau levels of charged particles are populated for above values of central 
field. For a 
particular temperature, the hyperon bulk viscosity coefficient is reduced 
compared with that of the zero field case. 
Further it is noted that the hyperon bulk viscosity decreases with increasing 
temperature as was earlier reported for the field free case.
Bulk viscosity coefficients due to dUrca processes in
a magnetic field have contributions from the forbidden as well as allowed 
domains. The bulk viscosity coefficients in magnetic fields having
central values $B_c = 10^{16}$ and $10^{17}$ G are enhanced in the allowed 
domain at higher baryon densities than those of field free cases.
We find an inversion of the temperature dependence of dUrca bulk viscosity
coefficients at $10^{11}$ K. We briefly discuss the effects of a superstrong 
magnetic field $\sim 10^{19}$ G on hyperon and dUrca bulk viscosities when
zeroth Landau levels of charged particles are populated. However, such a
superstrong magnetic field may not be a possibility in neutron stars.

In this calculation, we adopt the field free hyperon bulk viscosity relation
when protons populate large number of Landau levels. This may be an 
approximate treatment of the actual case. However the exact treatment of the 
effects of a magnetic field on the non-leptonic bulk viscosity 
would be worth studying when protons populate many Landau levels. Further 
the investigation of
bulk viscosity in magnetic fields has important implications for the r-modes in
magnetars. This will be reported in a future publication.


\newpage

\vspace{-2.0cm}

{\centerline{
\epsfxsize=12cm
\epsfysize=14cm
\epsffile{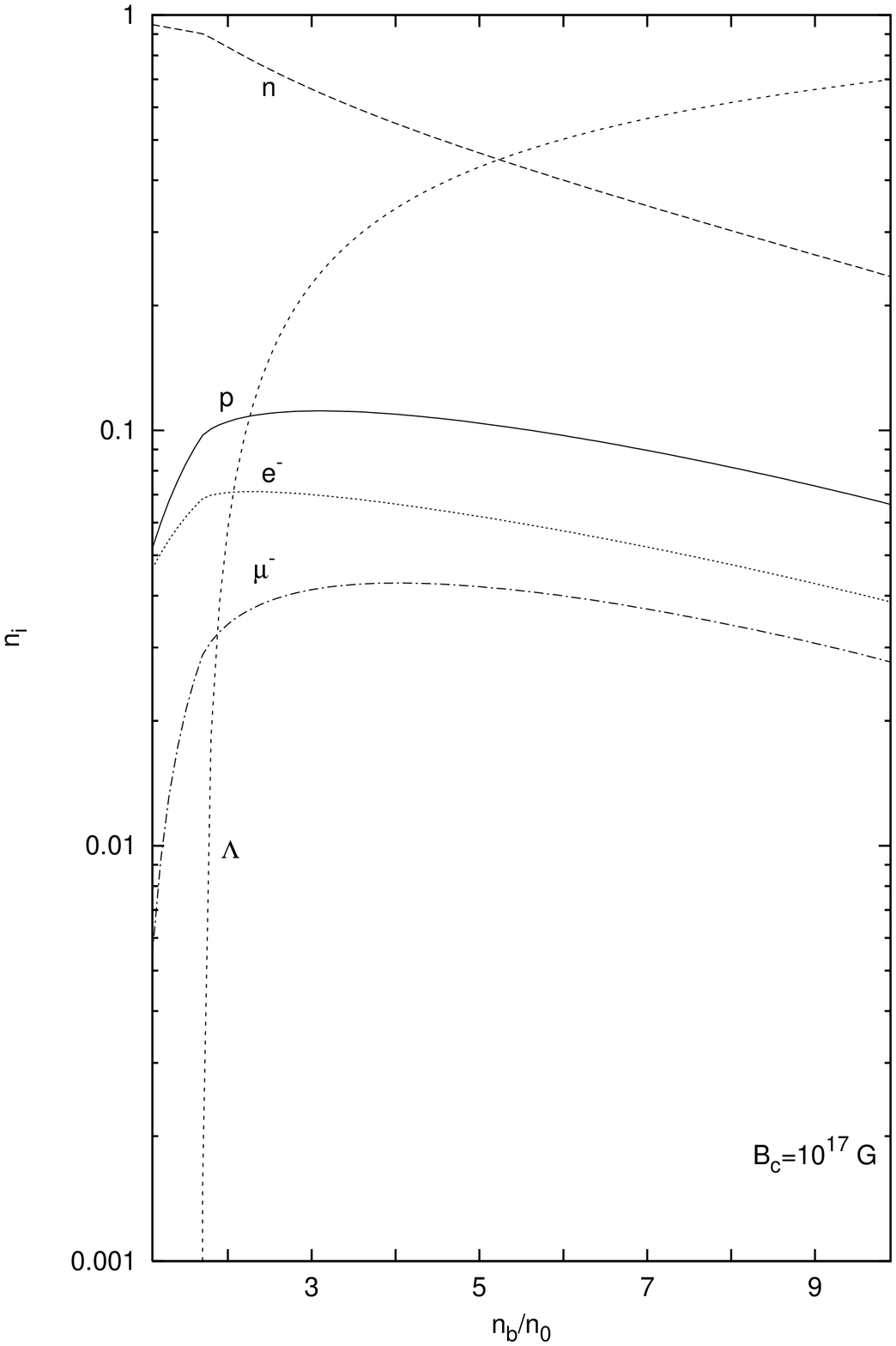}
}}

\vspace{1.0cm}

\noindent{\small{
Fig. 1. Fractions of different particle species 
in $\Lambda$-hyperon matter in the presence of a magnetic field having central
value $B_c = 10^{17}$ G as a function of normalised baryon density.}}
\label{fig:fraction}

\newpage

\vspace{-2.0cm}

{\centerline{
\epsfxsize=12cm
\epsfysize=14cm
\epsffile{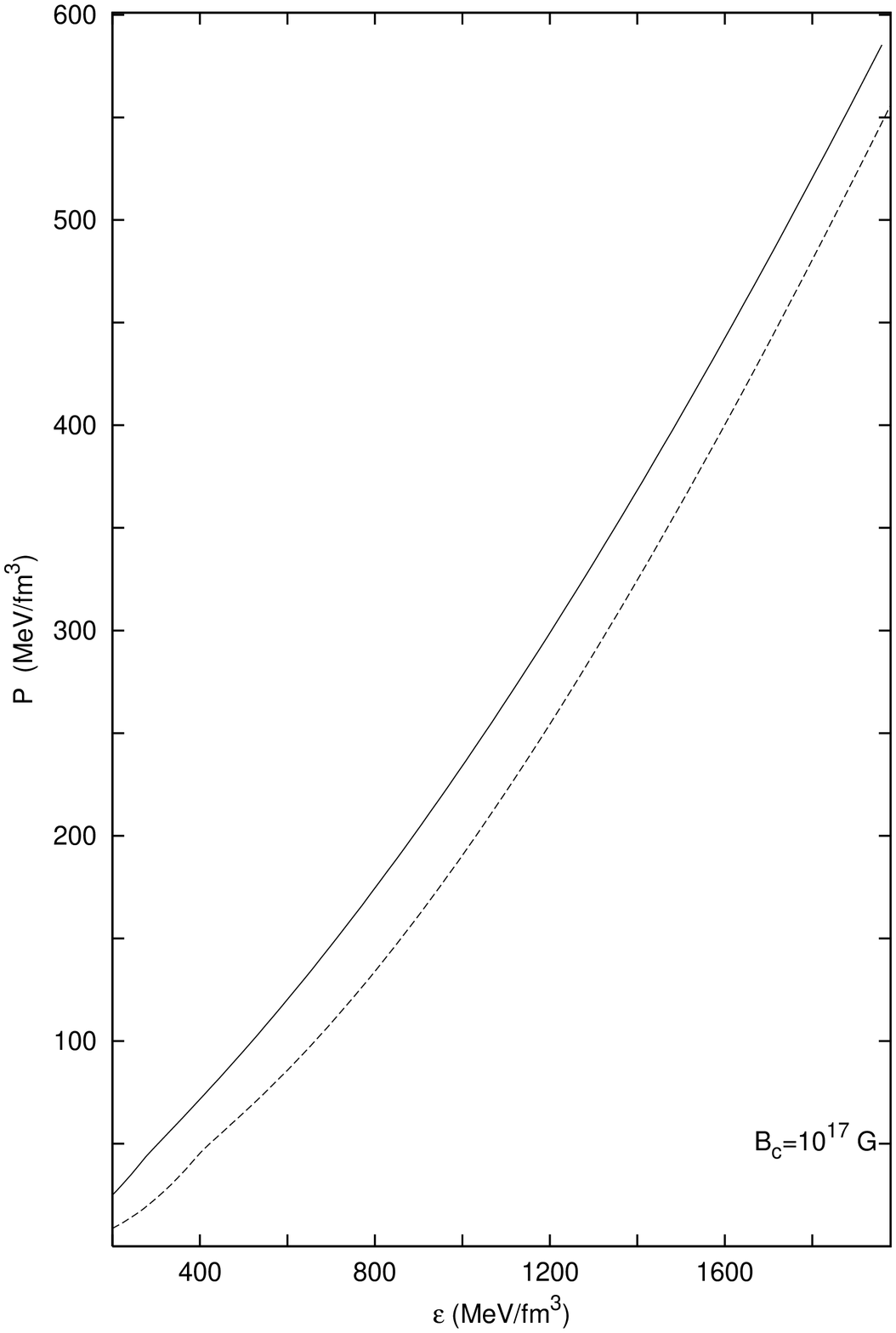}
}}

\vspace{1.0cm}

\noindent{\small{
Fig. 2. 
Equation of state, pressure versus energy density, with a magnetic field having
central value $B_c = 10^{17}$ G (solid line) and without magnetic field
(dashed curve).}} 

\newpage

\vspace{-2.0cm}

{\centerline{
\epsfxsize=12cm
\epsfysize=14cm
\epsffile{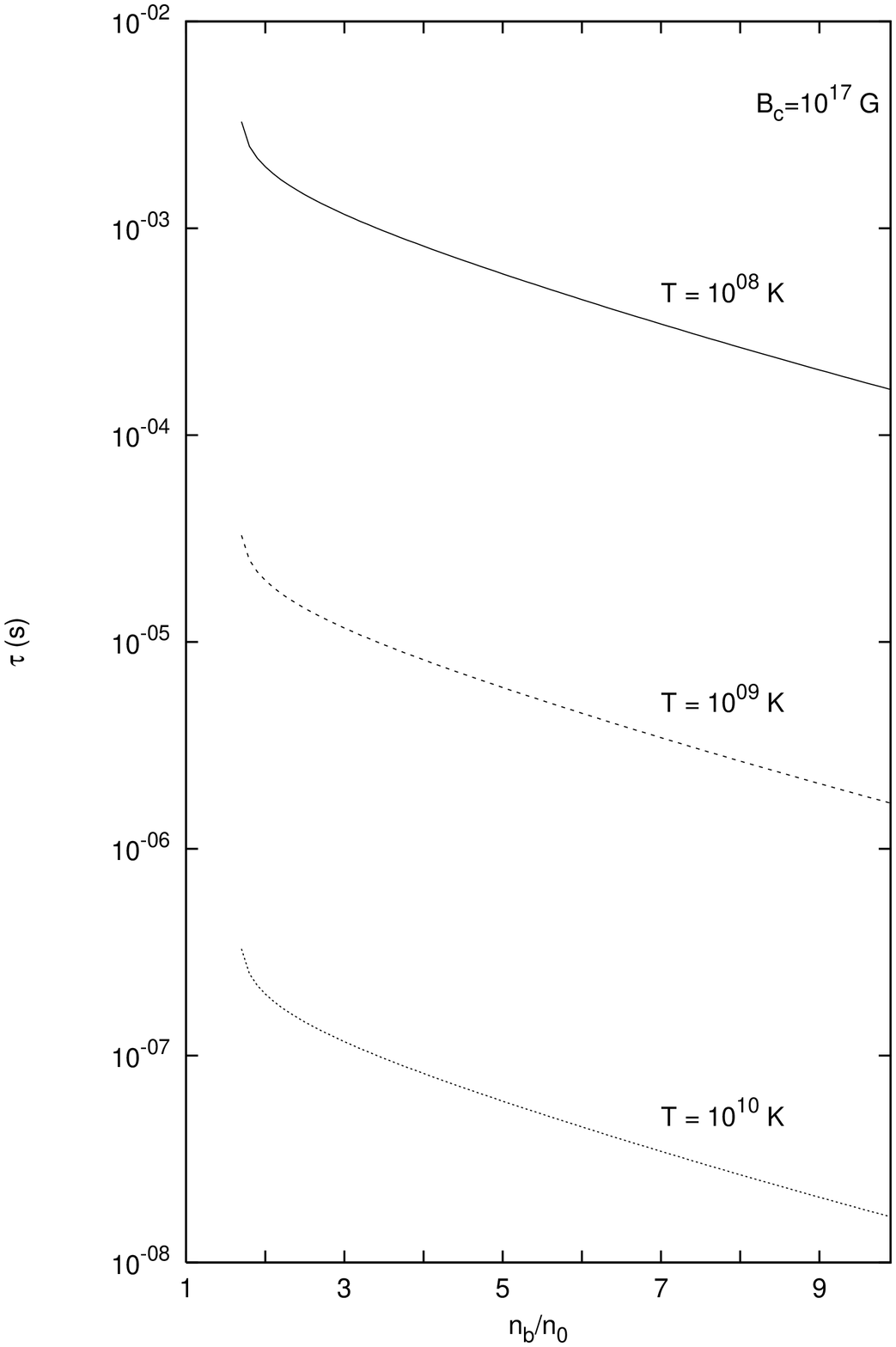}
}}

\vspace{1.0cm}

\noindent{\small{Fig. 3. Relaxation time for the non-leptonic reaction involving
$\Lambda$ hyperons in a magnetic field having central value $B_c = 10^{17}$ G 
and at different temperatures as a function of normalised baryon density.}} 

\newpage

\vspace{-2.0cm}

{\centerline{
\epsfxsize=12cm
\epsfysize=14cm
\epsffile{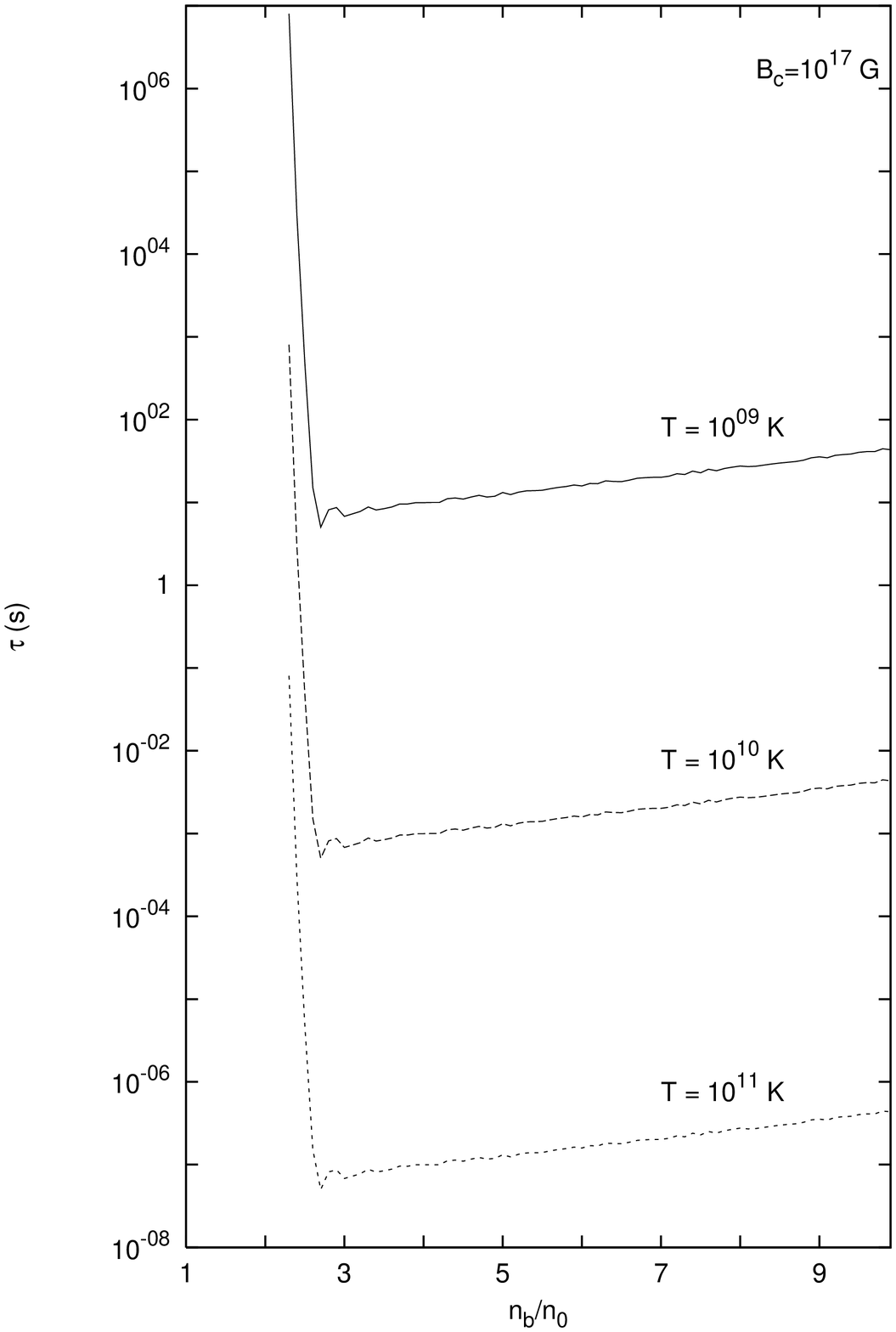}
}}

\vspace{1.0cm}

\noindent{\small{Fig. 4. Relaxation time of dUrca reaction involving electrons 
in a magnetic field having central value $B_c = 10^{17}$ G and at different 
temperatures as a function of normalised baryon density.}}
\label{fig:reactauu}

\newpage

\vspace{-2.0cm}

{\centerline{
\epsfxsize=12cm
\epsfysize=14cm
\epsffile{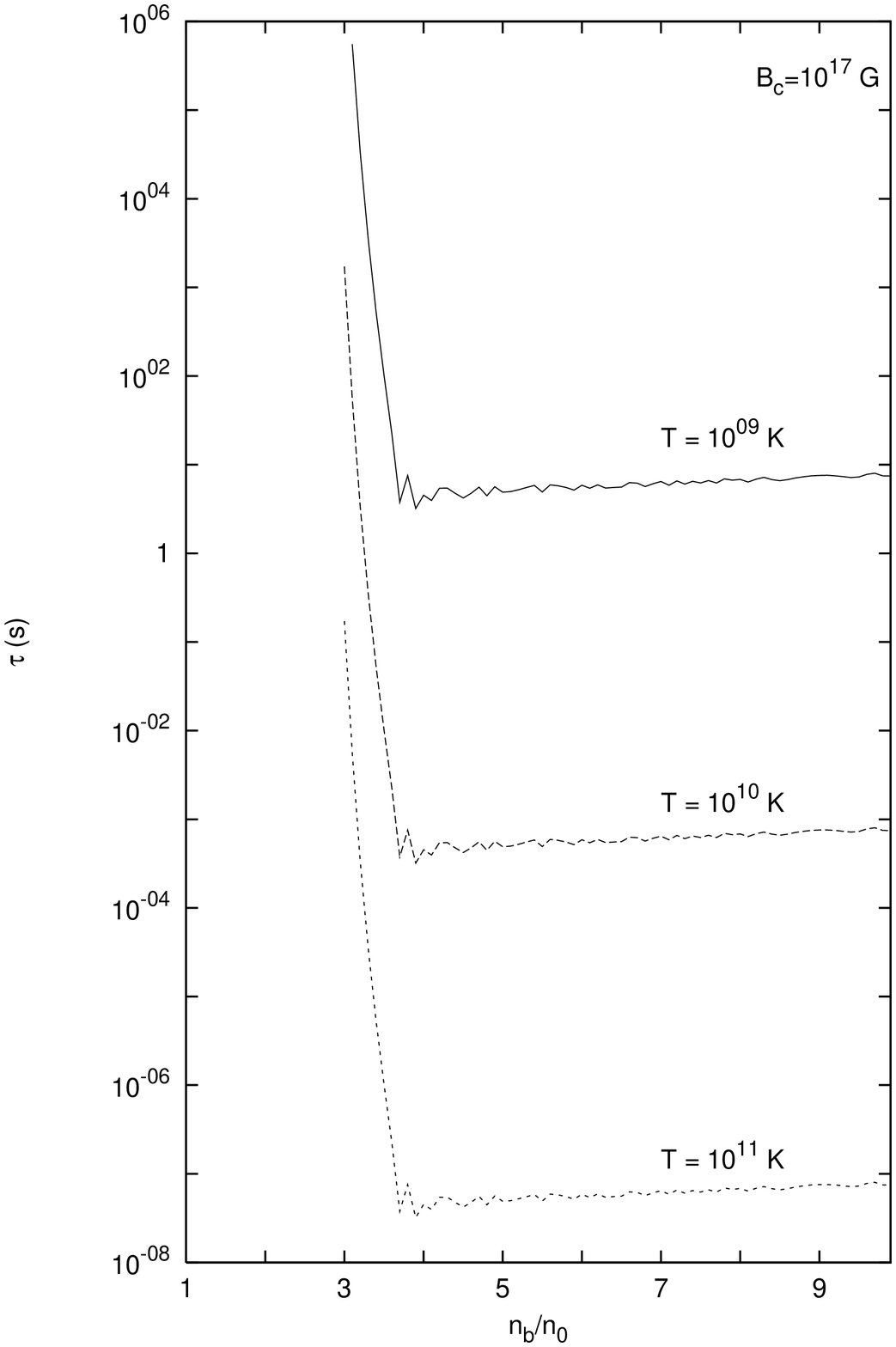}
}}

\vspace{1.0cm}

\noindent{\small{Fig. 5. same as Fig. 4 but for dUrca reaction including 
muons.}}

\newpage

\vspace{-2.0cm}

{\centerline{
\epsfxsize=12cm
\epsfysize=14cm
\epsffile{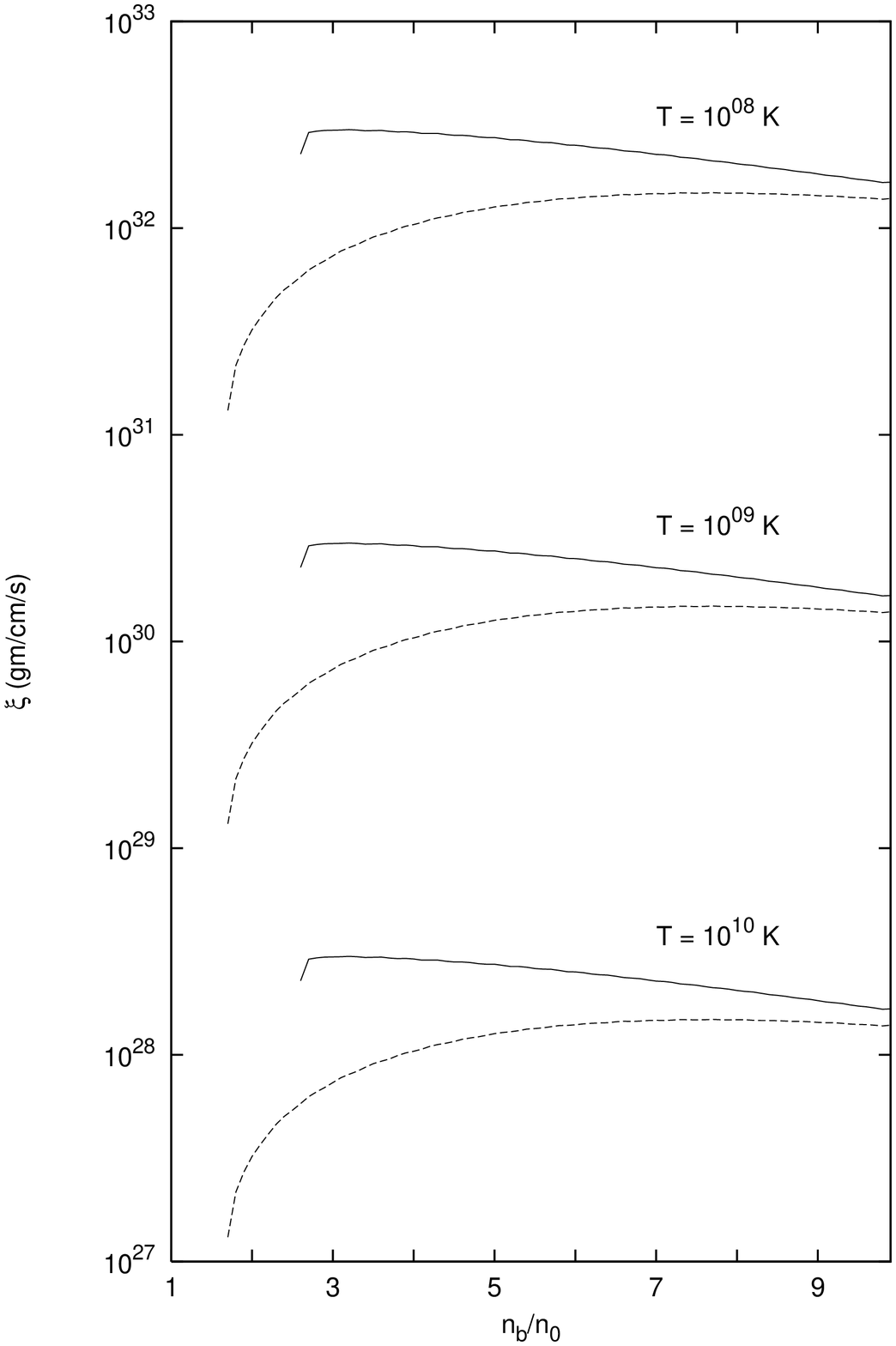}
}}

\vspace{1.0cm}

\noindent{\small{Fig. 6. Bulk viscosity coefficient (dashed line) for the 
non-leptonic processes involving $\Lambda$ hyperons in a magnetic field
having central value $B_c = 10^{17}$ G and at different temperatures as a 
function of normalised baryon density. Field free cases are shown by solid 
lines.}}
\label{fig:zeta}

\newpage

\vspace{-2.0cm}

{\centerline{
\epsfxsize=12cm
\epsfysize=14cm
\epsffile{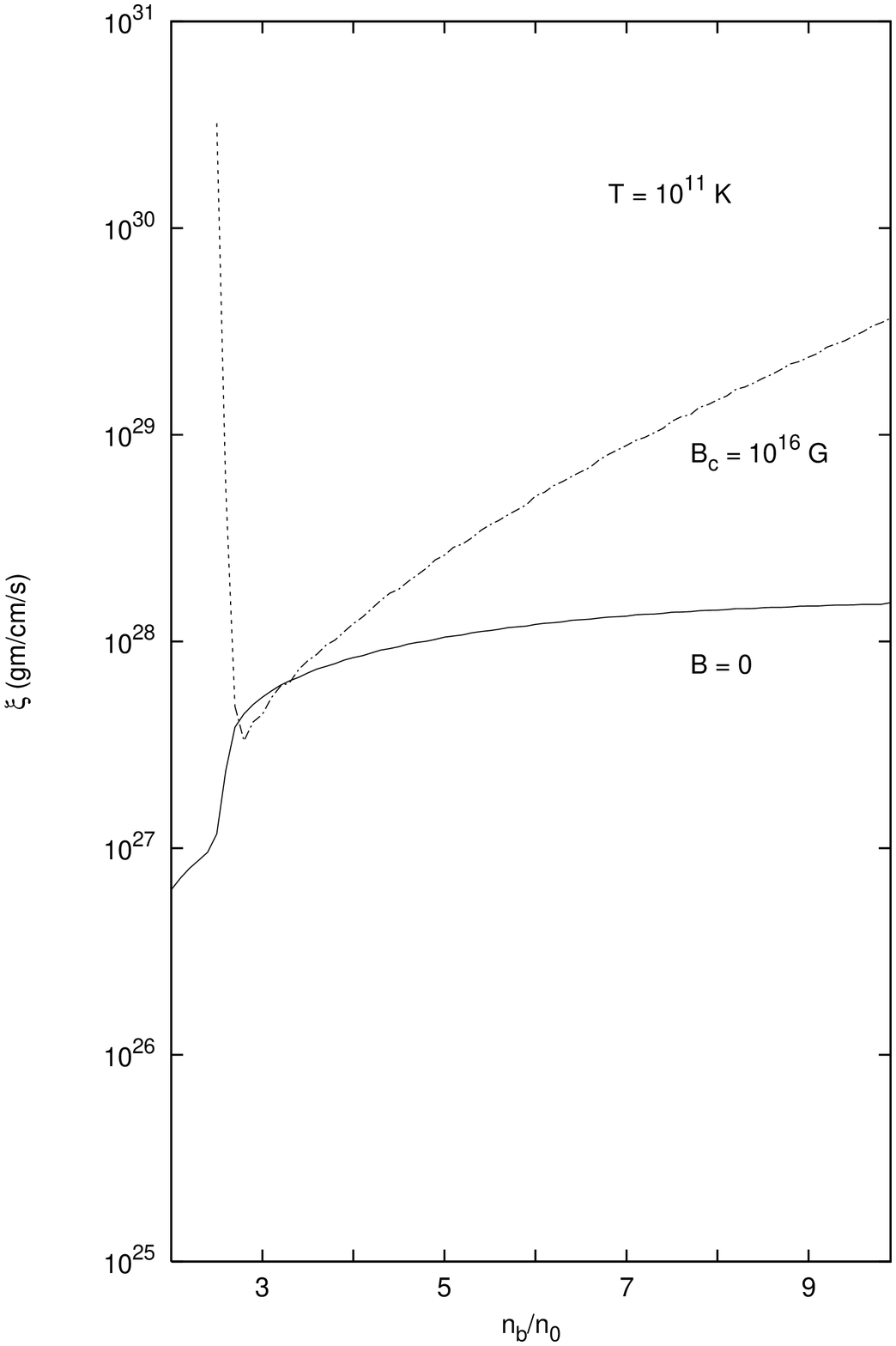}
}}

\vspace{1.0cm}

\noindent{\small{Fig. 7. Bulk viscosity coefficient for the dUrca process 
involving electrons in a magnetic field having central value $B_c = 10^{16}$ G 
and at a temperature $10^{11}$ K as a function of normalised baryon density.
The field free case is shown by the solid line.}}

\newpage

\vspace{-2.0cm}

{\centerline{
\epsfxsize=12cm
\epsfysize=14cm
\epsffile{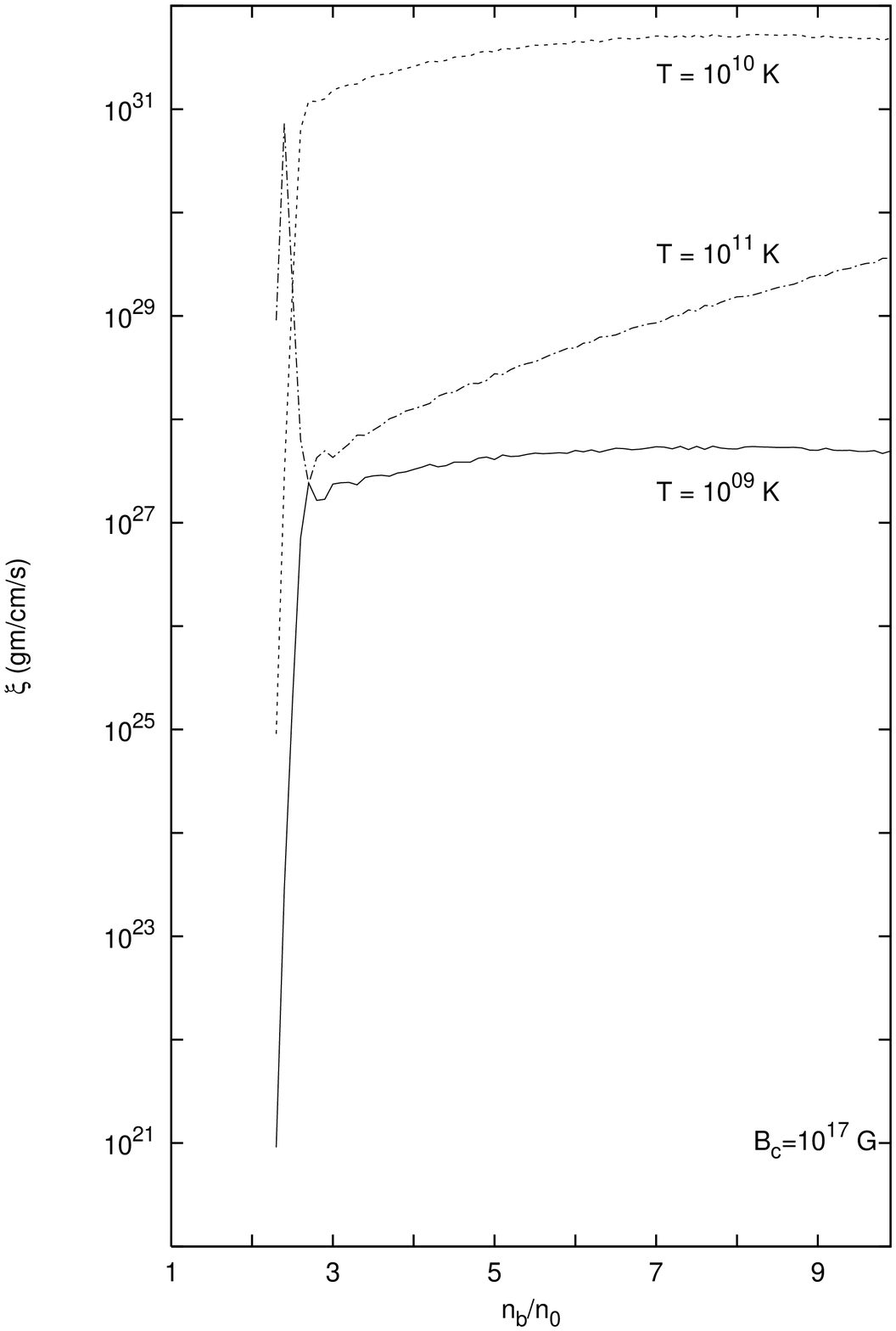}
}}

\vspace{1.0cm}

\noindent{\small{Fig. 8. Bulk viscosity coefficient for the dUrca process 
involving electrons in a magnetic field having central value $B_c = 10^{17}$ G 
and at different temperatures as a function of normalised baryon density.}}

\newpage

\vspace{-2.0cm}

{\centerline{
\epsfxsize=12cm
\epsfysize=14cm
\epsffile{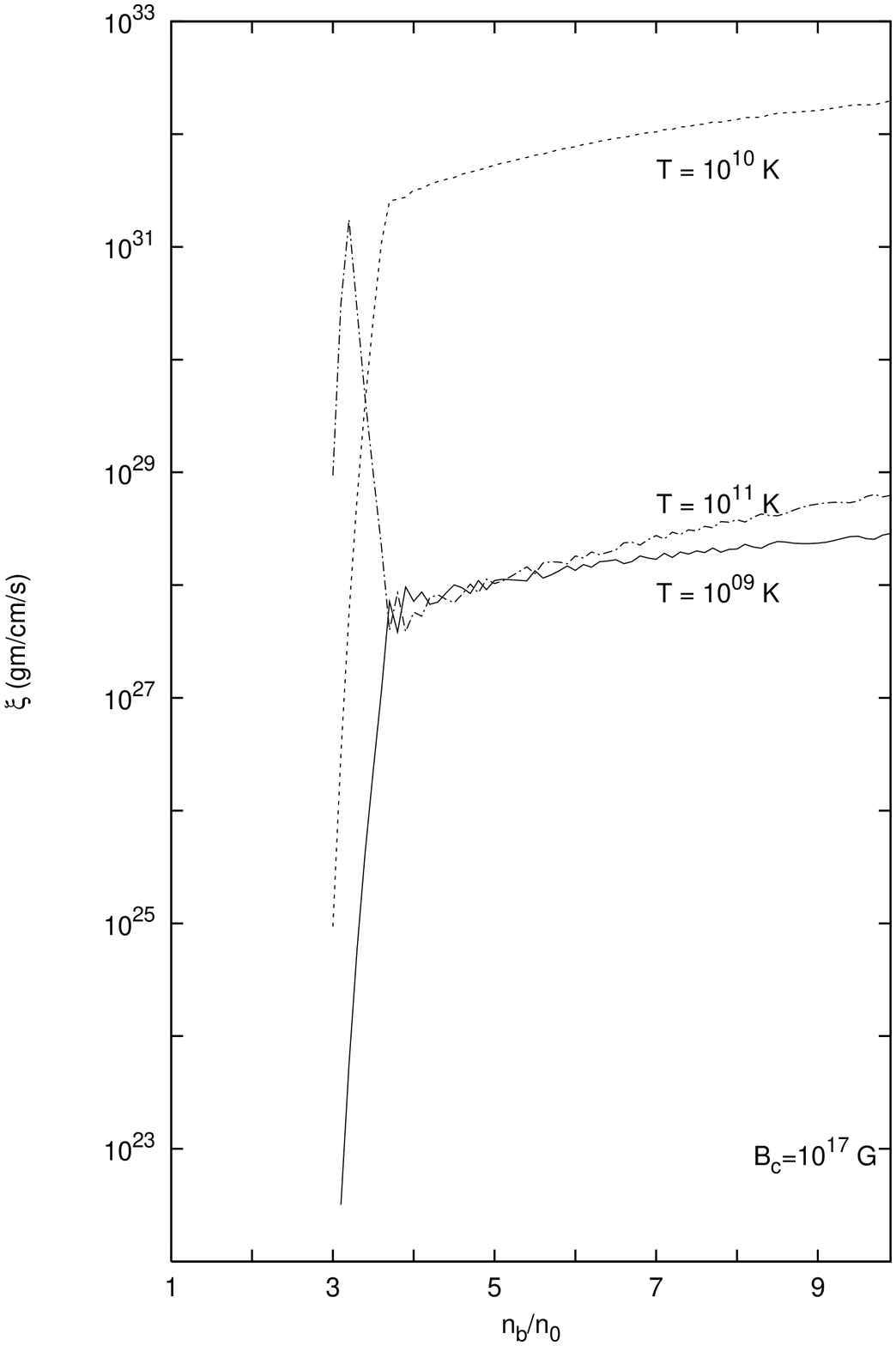}
}}

\vspace{1.0cm}

\noindent{\small{Fig. 9. Same as Fig. 8 but for the dUrca process involving
muons.}}

\end{document}